\documentclass[11pt]{article}
\usepackage{fullpage}
\usepackage{epsfig}
\usepackage{subfigure}
\usepackage{calc}
\usepackage{amssymb}
\usepackage{url}
\usepackage{amstext}
\usepackage{amsmath}
\usepackage{multicol}
\usepackage{pslatex}
\usepackage{ifpdf}

\newtheorem{theorem}{Theorem}

\begin{document}

\title{A note on quantum related-key attacks}


\author{Martin R{\"o}tteler\footnote{This work was carried out while MR was with NEC Laboratories America, Inc., Princeton, NJ 08540, USA.}\\
Microsoft Research\\
One Microsoft Way\\
Redmond, WA 98052, U.S.A.\\
{\tt martinro@microsoft.com}
\and Rainer Steinwandt\\
Florida Atlantic University\\
Department of Mathematical Sciences\\
Boca Raton, FL 33431\\
{\tt rsteinwa@fau.edu}
}

\maketitle

\begin{abstract}
In a basic related-key attack against a block cipher, the adversary has access to encryptions under keys that differ from the target key by bit-flips. In this short note we show that for a quantum adversary such attacks are quite powerful: if the secret key is (i) uniquely determined by a small number of plaintext-ciphertext pairs, (ii) the block cipher can be evaluated efficiently, and (iii) a superposition of related keys can be queried, then the key can be extracted efficiently.
\end{abstract}

\section{\uppercase{Introduction}}
\label{sec:introduction}
The availability of scalable quantum computers would jeopardize the security of many currently deployed asymmetric cryptographic schemes \cite{Sho97}. For symmetric cryptography the expectations for a post-quantum setting tend to be more optimistic, 
see e.g. \cite{Ber09}, from which we quote \emph{``quantum computers seem to have very little effect on secret-key cryptography, hash functions, etc. Grover's algorithm forces somewhat larger key sizes for secret-key ciphers, but this effect is essentially uniform across ciphers; today's fastest pre-quantum 256-bit ciphers are also the fastest candidates for post-quantum ciphers at a reasonable security level.''}

Related-key attacks are a powerful cryptanalytic tool when exploring block ciphers. In such attacks, the adversary is granted access to encryptions and/or decryptions of messages under secret keys which are related to the target key in a known or chosen way. As argued in \cite{Kel96}, this type of attack is of practical interest, despite the assumptions made. When Winternitz and Hellman described this attack model more than 25 years ago, they focused on key relations given by bit-flips \cite{WinHel87}. An illustrative example for an application of this attack model is an attack against 9 rounds of Rijndael with a 256-bit key, invoking 256 related keys with a particular choice of the bit-flips \cite{Ferg00}. 

Current approaches to formalize related-key attacks allow more general key relations \cite{BeKo06,AFPW11}, and restricting to bit-flips can be considered to be a rather conservative choice. 
Below we show that for a quantum adversary such a basic form of related-key attack is quite powerful. We show that the possibility to query a superposition of related keys to a block cipher enables the efficient extraction of the secret key, if some rather mild conditions are met:
\begin{enumerate}
  \item the block cipher can be implemented efficiently as a quantum circuit, and
  \item the secret key is uniquely determined by a small number of available plaintext-ciphertext pairs.
\end{enumerate}
The attack we describe is unlikely to pose a practical threat as querying a superposition of secret keys may not be feasible for a typical implementation. Notwithstanding this, from the structural point of view our observation indicates an interesting limitation for the security guarantees of a block cipher that one can hope to prove in a post-quantum scenario.

\section{\uppercase{Preliminaries}}\label{sec:prelims}

A block cipher with \emph{key length} $k$ and \emph{block length} $n$ is a family of $2^k$ permutations $\{E_K:\ \{0,1\}^n\longrightarrow\{0,1\}^n\}_{K\in\{0,1\}^k}$
on bitstrings of length $n$.
Popular block ciphers limit the possible choices of the key length $k$---e.\,g., for the Advanced Encryption Standard \cite{FIPS197} we have $n=128$ and $k\in\{128, 192, 256\}$. To characterize the efficiency of certain types of attacks, it can nonetheless be convenient to consider families of block ciphers, interpreting the key length $k$ as a scalable security parameter. Measuring the running time of an adversary as a function of $k$, it is meaningful to speak of an expected polynomial time attack.

\subsection{Related-key attacks}\label{sec:attackmodel}
The attack model we consider goes back to \cite{WinHel87}. After a key $K\in\{0,1\}^k$ has been chosen uniformly at random, the adversary has access to two oracles:
\begin{description}
\item[${\mathcal E}$:] On input a bitmask $L\in\{0,1\}^k$ and a bitstring $m\in\{0,1\}^n$, this oracle returns the encryption $E_{K\oplus L}(m)$ of $m$ under the key $K\oplus L$.
\item[${\mathcal E^{-1}}$:] On input a bitmask $L\in\{0,1\}^k$ and a bitstring $c\in\{0,1\}^n$, this oracle returns the decryption $E_{K\oplus L}^{-1}(c)$ of $c$ under the key $K\oplus L$.
\end{description}
After interacting with these oracles, the adversary has to output a guess $K'$ for $K$, and it is considered successful if and only if $K=K'$. For our attack we will also assume that the block cipher at hand can be evaluated efficiently, i.\,e., with a polynomial-size quantum circuit that has the secret key and a plaintext as input. For block ciphers that are actually used this condition is of no concern.

The quantum attack below will not involve ${\mathcal E}^{-1}$, but we will allow the adversary to query the block cipher and also the oracle $\mathcal E$ with a superposition of keys. Finally, we require that the adversary has access to a polynomial number of plaintext-ciphertext pairs $(m_1,c_1),\dots,(m_r,c_r)$ such that there exists exactly one secret key $K\in\{0,1\}^k$ satisfying
$$(c_1,\dots,c_r)=(E_K(m_1),\dots,E_K(m_r)).$$
It is easy to come up with a pathological block cipher where the secret key cannot be uniquely determined by any number of plaintext-ciphertext pairs\footnote{Encryption and decryption can simply ignore parts of the secret key.}, but for typical block ciphers we do not think this to be a concern. In \cite[Definition~7.34]{HAC01} the \emph{known plaintext unicity distance} is defined as a measure for the number of (known) plaintext-ciphertext pairs that are needed to determine the secret key of a block cipher uniquely, and with \cite[Fact~7.35]{HAC01} it seems plausible to estimate that for an $n$-bit block cipher with key length $k$ having 
\begin{equation}
r>\lceil k/n\rceil\label{equ:kpunicity}
\end{equation}
plaintext-ciphertext pairs suffices. So for the 128-bit version of AES, where $n=k=128$, one can think of an $r$-value as small as $2$. Throughout we will assume that $r$ satisfies Inequality~\eqref{equ:kpunicity}. Then the main idea to mount a quantum related-key attack is a reduction to a quantum algorithm described in \cite{Sim94} which we describe next.

\subsection{Simon's problem}

Let $f:\ \{0,1\}^k\longrightarrow \{0,1\}^{k'}$ with $k\le k'$ be a function such that one of the following two conditions holds:
\begin{description}
\item[(a)] $f$ is injective;
\item[(b)] there exists a bitstring $s\in \{0,1\}^k\setminus\{0^k\}$ such that for every two distinct $x,x'\in\{0,1\}^k$ we have $$f(x)=f(x')\iff x=x'\oplus s.$$
\end{description}
Simon's problem asks to decide for such a function $f$ which of the two conditions holds, and in the case (b) to find $s$. Allowing the function $f$ to be evaluated at a superposition of inputs, \cite{Sim94} establishes the following result:
\begin{theorem}\label{the:simon}
Let $g(k)$ be an upper bound for the time needed to solve a $k\times k$ linear system of equations over the binary field ${\mathbb F}_2$, and let $t_f(k)$ be an upper bound for the time needed to evaluate the function $f$ on (a superposition of) inputs from $\{0,1\}^k$.
Then the above problem can be solved in expected time $\mathrm{O}(k\cdot t_f(k)+g(k))$. In particular, for $t_f=t_f(k)$ being polynomial, the above problem can be solved in expected polynomial time.
\end{theorem}

\begin{figure*}[hbt]
\centerline{
\includegraphics[width=\textwidth]{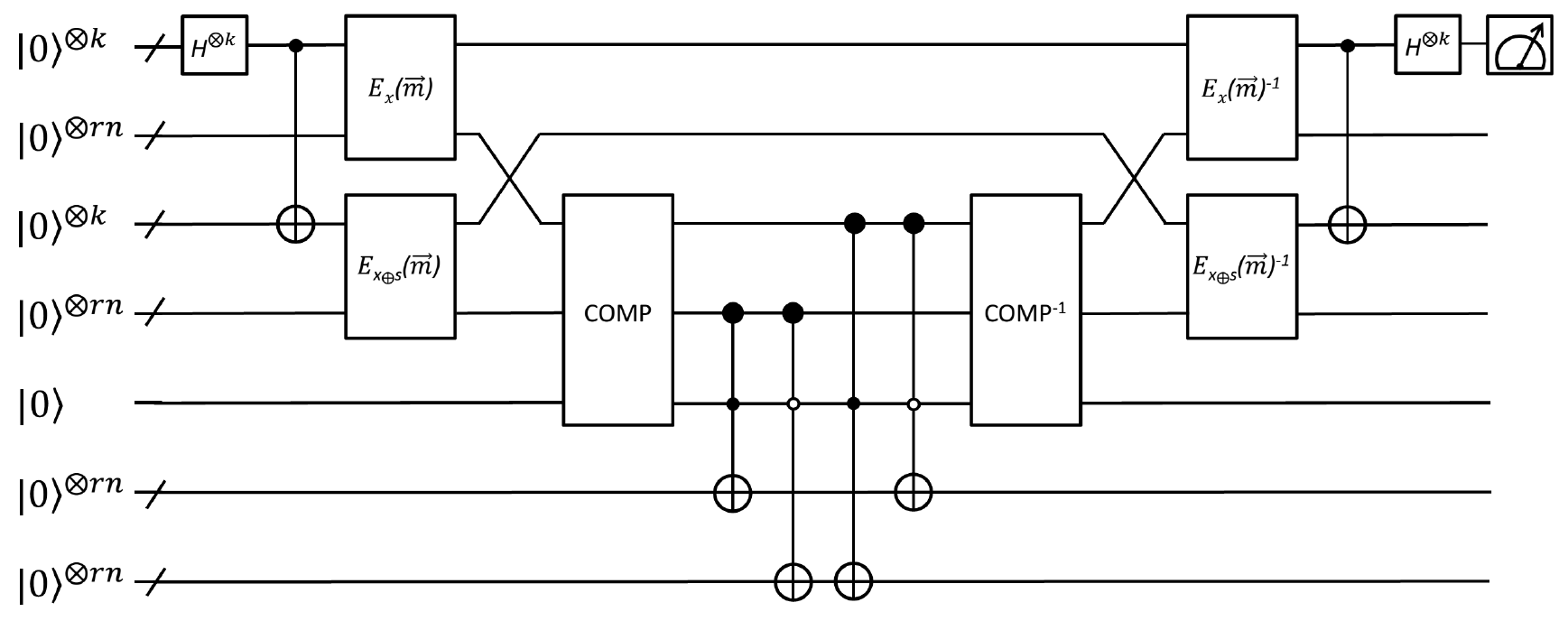}
}
\caption{\label{fig:RKcircuit} Quantum circuit to implement a quantum related-key attack on a block cipher via a reduction to an instance of Simon's problem. The basic building blocks of this circuit are described in more detail in the text.  The circuit makes use of calls to the block cipher for keys $x$ to obtain encryptions $E_x(\vec{m})$ and of calls to the related-key oracle to obtain the corresponding encryptions $E_{x \oplus s}(\vec{m})$ with respect to the related keys $x \oplus s$ that are obtained from the secret key $s$ via exclusive-OR (XOR) masks. Further detail on the comparison circuits is given in Figure \ref{fig:comp}. In the center of the circuit is a sequence of four copy operations via CNOTs to pick out the desired register and copy it into the target registers holding $\min(c,c^\prime)$ and $\max(c,c^\prime)$, respectively. We make use of a graphical notation that employs control knobs of different sizes which is described further in Figure \ref{fig:copy}. Together, the comparison and copy operations realize the data structure of an unordered set which we use in our reduction to Simon's problem. Upon measurement of the first register, a $k$-bit vector $y \in \{0,1\}^k$ is obtained which due to our construction is perpendicular to the secret key $s \in \{0,1\}^k$. After ${\mathrm O}(k)$ iterations, with constant probability the secret key can be reconstructed from the measurement data.}
\end{figure*}

\section{Description of the attack}
Alluding to the Electronic Code Book mode of operation \cite[Section~7.2.2]{HAC01}, subsequently we will simply write ${E}_K(\vec m)$ for the tuple of ciphertext blocks $({E}_K(m_1), \dots, {E}_{K}(m_r))\in\{0,1\}^{rn}$. For a fixed, unknown secret key $s\in\{0,1\}^k\setminus\{0^k\}$ and messages ${\vec m}\in \{0,1\}^{rn}$ that characterize the $s$ uniquely as described in Section \ref{sec:prelims}, we define the function

$$\begin{array}{lccc}
  f_s:&\{0,1\}^k&\longrightarrow&2^{\{0,1\}^{rn}}\\
&x&\longmapsto&\{{E}_x(\vec m), {E}_{s\oplus x}(\vec m)\}
\end{array}.$$

We remark that for each $x$ in the domain of $f_s$, the image is comprised of two different ciphertexts, i.\,e., it does not collapse to a singleton set. Indeed, this is the case due to the choice of the plaintexts $m_1,\dots,m_r$ as the condition $s\ne 0^k$ implies that ${E}_x(\vec m)\ne {E}_{s\oplus x}(\vec m)$. We next describe our core result, namely a reduction from the problem of finding the secret key $s$ to an instance of Simon's problem which can then be solved efficiently on a quantum computer. 

\paragraph{Meeting the conditions of Simon's problem}
To argue that $f_s$ meets the conditions of Theorem~\ref{the:simon}, let us first clarify how to encode the images as elements of $\{0,1\}^{k'}$ for some $k'\ge k$. As we impose condition~\eqref{equ:kpunicity}, with $k'{=}(rn+rn){=}2rn$ we clearly have $k'\ge k$ as desired. By interpreting elements in $\{0,1\}^{rn}$ as (unsigned) integers, we can impose a linear order on $\{0,1\}^{rn}$. Then, to store an element $\{c, c'\}$ in the image of $f_s$, we simply store the ordered pair $(\min(c,c'), \max(c,c'))$ as its unique $k'$-bit representation.

Next, consider two different $k$-bit strings $x\ne x'$ that satisfy $f_s(x)=f_s(x')$:
\begin{itemize}
   \item If ${E}_{x}(\vec m)={E}_{x'}(\vec m)$ then the choice of the plaintexts $m_1,\dots,m_r$ implies $x=x'$, so this cannot happen.
   \item If ${E}_{x}(\vec m)\ne {E}_{x'}(\vec m)$, then  ${E}_{x}(\vec m)={E}_{s\oplus{x'}}(\vec m)$, which by the choice of the plaintexts $m_1,\dots, m_r$ means that $x=s\oplus x'$.
\end{itemize}
So we have the implication $f_s(x)=f_s(x')\Longrightarrow x=x'\oplus s.$
The converse follows trivially from $s\oplus(x'\oplus s)=x'$.

Next, let us check that the function $f_s(\cdot)$ can be evaluated efficiently.

\paragraph{Evaluating $f_s(\cdot)$ in polynomial time} By assumption the underlying block cipher can be evaluated with a polynomial-size quantum circuit, so computing the two values ${E}_{x}(\vec m)$ and ${E}_{s\oplus x}(\vec m)$ for a given $x$ can certainly be done in polynomial time. In the actual attack, the value  ${E}_{s\oplus x}(\vec m)$ is obtained by invoking the encryption oracle $\mathcal E$ for each of the plaintexts $m_1,\dots,m_r$, i.\,e., with a polynomial number of queries to $\mathcal E$. This means we can obtain the pair $({E}_{x}(\vec m), {E}_{s\oplus x}(\vec m))$ in polynomial time, and we are left to distill our unique $k'$-bit representation of the \emph{set} comprised by these two elements.

As indicated in the previous paragraph, such a representation can be implemented by interpreting the two ciphertexts as integers and by then sorting them. A quantum circuit to determine this unique representation of a pair of bitvectors consists of swapping ${E}_{x}(\vec m)$ and ${E}_{s\oplus x}(\vec m)$ conditioned on the the latter value being smaller than the former. For instance with a reversible circuit to perform addition \cite{CDKM04,DKRS06,TTK10} one can compute the difference of the binary numbers represented by ${E}_{x}(\vec m)$ and ${E}_{s\oplus x}(\vec m)$ in polynomial time. The most significant bit of the result then reveals the result of the comparison. The swap operation can be conditioned on this bit, followed by an uncomputation of the garbage introduced by the adder \cite{Bennett73}. 

\begin{figure}[hbt]
\centerline{
\qquad \includegraphics[width=0.33\columnwidth]{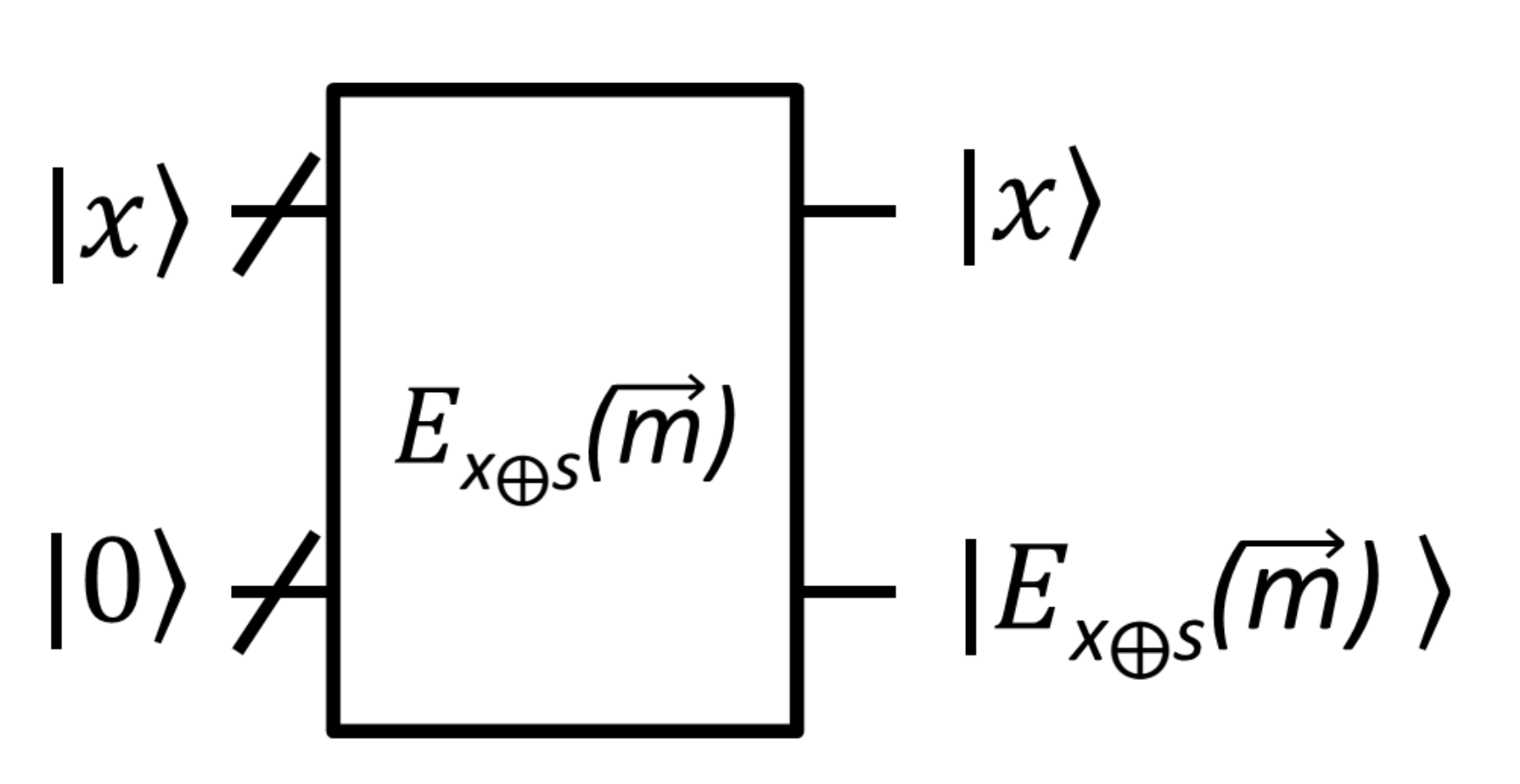}
}
\caption{\label{fig:oracle} Quantum oracle to implement the encryption of a tuple $\vec{m}$ of messages under an encryption key $x \oplus s$ that is related to the secret key $s$ via addition of an XOR mask $x$. Note that the circuit for $s=0$ is needed in Figure \ref{fig:RKcircuit} also and this circuit can---by assumption on the cipher---be efficiently implemented. The shifted version of the cipher has to be given as a block-box circuit which we can evaluate in superposition.} 
\end{figure}

\begin{figure}[hbt]
\centerline{
\includegraphics[width=0.33\columnwidth]{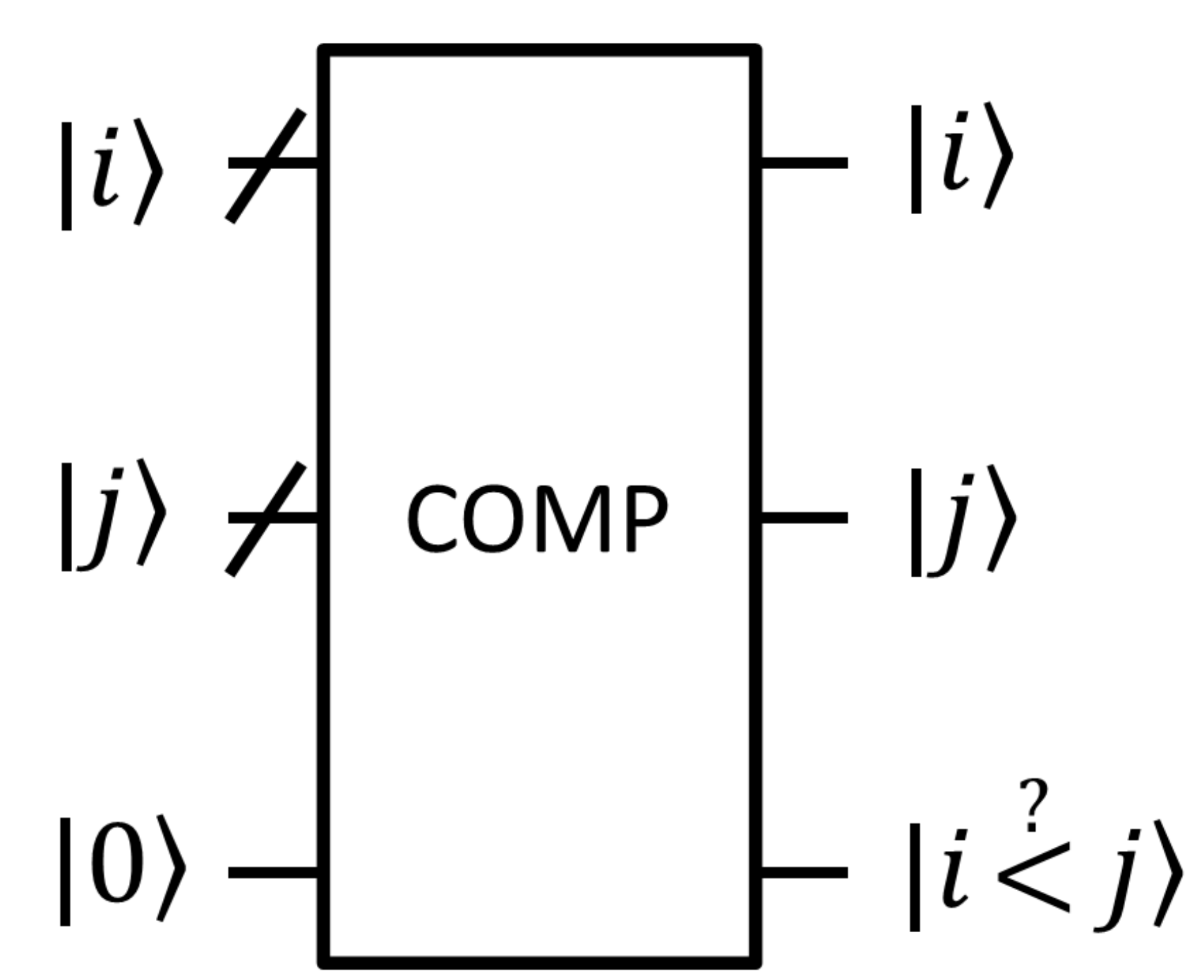}
}
\caption{\label{fig:comp} Quantum circuit for the comparison of two $n$ bit integers. An efficient implementation can, e.\,g., be obtained by computing the difference $i-j$ in one's complement form and then keeping only the highest order bit (see \cite{CDKM04,DKRS06}).}
\end{figure}

\begin{figure}[hbt]
\centerline{
\includegraphics[width=0.5\columnwidth]{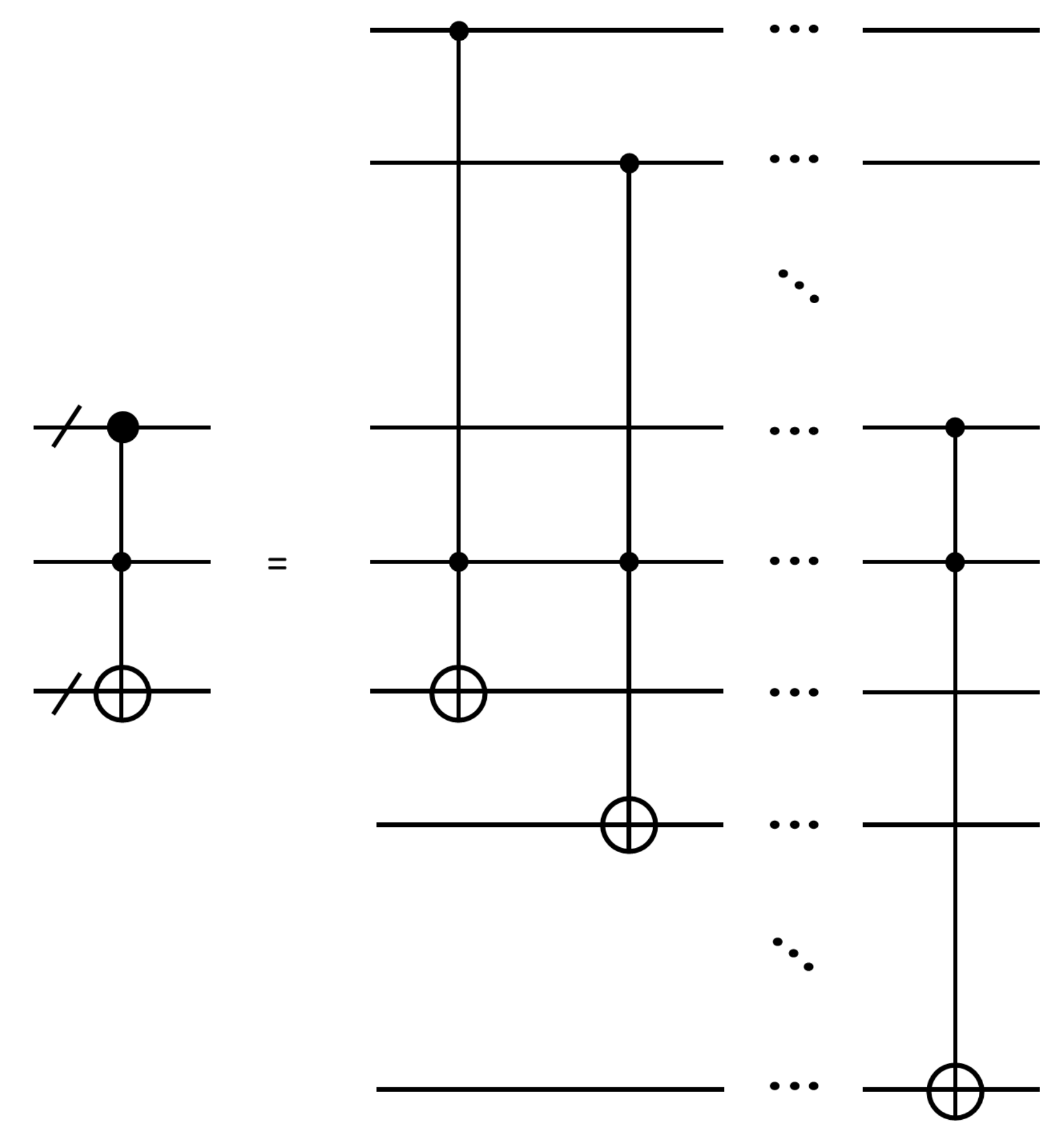}
}
\caption{\label{fig:copy} Quantum circuit to implement a controlled copy operation from one quantum register to another quantum register. In the figure, the source register consists of the upper $n$ qubits, the target register of the lower $n$ qubits, and the control qubit sits in the middle.} 
\end{figure}

In Figure \ref{fig:RKcircuit} we display the resulting quantum circuit. Note that each computation of an intermediate result has to be uncomputed or else interference between the various computational paths could not take place. 
The overall structure of the circuit is that of a Fourier sampling circuit, i.e., a circuit type that arises in the solution of abelian hidden subgroup problems \cite{BH:97,ME:98}. 

The implementation of the hiding function $f_s$ is decomposed into several subroutines to make the circuit more readable: in a first stage, the classical $k$-bit vector $x$ is fanned out into two copies using a CNOT gate (note that a line with a tick denotes a quantum register holding two or more qubits). The value $x$ is then passed to a subroutine computing $E_x(\vec{m})$, respectively $E_{x \oplus s}(\vec{m})$ in superposition. These subroutines are carried out by circuits as in Figure \ref{fig:oracle}. An efficient cirucit for $E_x(\vec{m})$ can be synthesized by making the efficient circuit (which by assumption exists) reversible. For the implementation of $E_{x \oplus s}(\vec{m})$ we make use of the related-key oracle ${\cal E}$. Throughout, it should be noted that $\vec{m}$ is a vector that uniquely characterizes $s$ as in Section \ref{sec:attackmodel}.  

The ``COMP'' operation is shown in Figure \ref{fig:comp} and can be realized similarly to addition of integers. See also \cite{CDKM04,DKRS06} for implementations of comparison circuits that optimize    
circuit width, respectively circuit depth. The result of the comparison is then used to copy the smaller of the two registers (when interpreted as integers) into the uppermost of the last two registers and the larger one into the lower one. Finally, in Figure \ref{fig:copy} one of the operations is shown that 
allow to select one of the two registers holding an $rn$-bit integer, depending on the value of the comparison operation. The other type which picks a register controlled by the negated value of the comparison bit is implemented analogously. Overall we have established the following result. 

\begin{theorem}\label{the:prop}
For every $s\in\{0,1\}^k\setminus\{0^k\}$ the function~$f_s$ defined above satisfies the conditions needed to apply Theorem~\ref{the:simon}, and the bound $t_{f_s}$ can be chosen to be polynomial.
\end{theorem}

By combining Theorems \ref{the:simon} and \ref{the:prop} we now obtain the following quantum related-key attack which runs in expected polynomial time:

\begin{enumerate}
  \item Check if the secret target key $s$ is the all-zero key $s=0^k$ by computing ${E}_{0^k}(\vec m)$ and comparing these ciphertexts with the given ciphertexts ${E}_s(\vec m)$.
  \item If $s\ne 0^k$ then apply Simon's algorithm---which constitutes the proof of Theorem~\ref{the:simon}---to recover $s$.
\end{enumerate}

\section{Conclusion}
This note shows that in a quantum setting even a basic related-key attack is very powerful: under rather mild assumptions on the attacked block cipher the secret key can be extracted efficiently.

\section*{\uppercase{Acknowledgments}}
\noindent RS was supported by the Spanish \emph{Ministerio de Econom{\'\i}a y Competitividad} through the project grant MTM-2012-15167 and by NATO's Public Diplomacy Division in the framework of ``Science for Peace'', Project MD.SFPP 984520. This work was carried out while MR was with NEC Laboratories America, Princeton, NJ 08540, U.S.A. We thank Schloss Dagstuhl, Germany, for providing an excellent research environment in which part of this research was carried out during a {\em Quantum Cryptanalysis} seminar.



\end{document}